\documentstyle[12pt]{article}

\begin{document}

\title{Quasi-exact Solvability of Planar Dirac Electron\\
in Coulomb and Magnetic Fields}

\author{Chun-Ming Chiang$^{1}$ and Choon-Lin Ho$^2$}
\date
{\small \sl $^1$Northern Taiwan Institute of Science and
Technology, Peitou, Taipei 112, Taiwan, R.O.C.\\$^2$Department of
Physics, Tamkang University, Tamsui 25137, Taiwan, R.O.C.}

\maketitle

\begin{abstract}
The Dirac equation for an electron in two spatial dimensions in
the Coulomb and homogeneous magnetic fields is a physical example
of quasi-exactly solvable systems. This model, however,  does not
belong to the classes based on the algebra $sl(2)$ which underlies
most one-dimensional and effectively one-dimensional quasi-exactly
solvable systems. In this paper we demonstrate that the
quasi-exactly solvable differential equation possesses a hidden
$osp(2,2)$ superalgebra.
\end{abstract}



\date{Dec 13, 2004} 

\newpage

\section{Introduction.}

It is well known that exactly solvable models play an important
role in many fields of physics. However, exactly-solvable systems
are very rare in physics.
 Recently, in quantum mechanics a new type of spectral problems,
which is intermediate to exactly solvable ones and non solvable
ones, has been found [\cite{TU}-\cite{Tur3}].  For this new class
of spectral problems, the so-called quasi-exactly solvable (QES)
models, it  is possible to determine algebraically a part of the
spectrum but not the whole spectrum. Such quasi-exact solvability
is usually due to some hidden Lie-algebraic structures. More
precisely, a QES Hamiltonian can be reduced to a quadratic
combination of the generators of a Lie group with
finite-dimensional representations.

The first physical example of QES model is the system of two
electrons moving in an external oscillator potential discussed in
\cite{SG,Taut1}.  Later, several physical QES models were
discovered, which include the two-dimensional Schr\"odinger
\cite{Taut2}, the Klein-Gordon \cite{VP}, and the Dirac equation
\cite{HoKha,ChHo1} of an electron moving in an
attractive/repulsive Coulomb field and a homogeneous magnetic
field.  More recently, the Pauli and the Dirac equation minimally
coupled to magnetic fields \cite{HoRoy1}, and Dirac equation of
neutral particles with non-minimal electromagnetic couplings
\cite{HoRoy2} were also shown to be QES.

It turns out that the system of two electrons moving in an
external oscillator potential and those of an electron moving in
an Coulomb field and a homogeneous magnetic field mentioned above,
with the exception of the Dirac case, share the same underlying
structure that made them QES. In \cite{ChHo2} it was demonstrated
that these systems are governed essentially by the same basic
equation, which is QES owing to the existence of a hidden $sl(2)$
algebraic structure.  This algebraic structure was first realized
by Turbiner for the case of two electrons in an oscillator
potential \cite{Tur4}. For the Dirac case, on the other hand, it
had been shown the quasi-exact solvability of this system is not
related to the $sl(2)$ algebra \cite{ChHo1}. At that time,
however, it was not known to us what algebraic structure the Dirac
system possesses.  In this paper we would like to show that the
relevant symmetry of this system is the $osp(2,2)$ superalgebra.

\vskip 0.5 truecm

\section{The Dirac Equation}

In $2+1$ dimension, the Dirac equation for an electron minimally
coupled to an external electromagnetic field has the form (we set
$c=\hbar=1$)
\begin{eqnarray}
(i\partial_t - H_D)\Psi(t, {\bf r}) = 0, \label{eq2}
\end{eqnarray}
where
\begin{eqnarray}
  H_D = \sigma_1P_2 -
\sigma_2P_1 + \sigma_3m - eA^0 \label{eq3}
\end{eqnarray}
is the Dirac Hamiltonian, $\sigma_k$ ($k=1,2,3$) are the Pauli
matrices, $P_{k} = -i\partial_k + eA_k$ is the operator of
generalized momentum of the electron, $A_{\mu}$ the vector
potential of the external electromagnetic field, $m$ the rest mass
of the electron, and $-e~ (e>0)$ is its electric charge. The Dirac
wave function $\Psi(t, {\bf r})$ is a two-component function. For
an external Coulomb field and a constant homogeneous magnetic
field $B>0$ along the $z$ direction, the potential $A_\mu$ are
given by
\begin{eqnarray}
 A^0(r) = Ze/r~ (e>0), \quad A_x = -By/2, \quad A_y = Bx/2
\label{e1}
\end{eqnarray}
in the symmetric gauge. The wave function is assumed to have the
form
\begin{eqnarray}
 \Psi(t,{\bf x}) = \frac{1}{\sqrt{r}}\exp(-iEt)
\psi_l(r, \varphi)~, \label{e3}
\end{eqnarray}
where  $E$ is the energy of the electron, and
\begin{eqnarray}
\psi_l(r, \varphi) = \left( \begin{array}{c}
F(r)e^{il\varphi}\\
G(r)e^{i(l+1)\varphi}
\end{array}\right)
\label{eqn6}
\end{eqnarray}
with integral number $l$. The function $\psi_l(r,\varphi)$ is an
eigenfunction of the conserved total angular momentum $J_z=L_z +
S_z = -i\partial/\partial\varphi + \sigma_3/2$ with eigenvalue
$j=l+1/2$.    It should be reminded that $l$ is not a good
quantum number.  Only the eigenvalues $j$ of the conserved total
angular momentum $J_z$ are physically meaningful.

Putting Eq.(\ref{e3}) and (\ref{eqn6}) into (\ref{eq2}), and
taking into account of the equations
\begin{eqnarray}
 P_x \pm iP_y = -ie^{\pm i\varphi}\left(\frac{\partial}{\partial r} \pm
\left(\frac{i}{r}\frac{\partial}{\partial\varphi}-\frac{eBr}{2}\right)
\right)~, \label{impul}
\end{eqnarray}
we obtain
\begin{eqnarray}
\frac{dF}{dr} - \left(\frac{l+\frac{1}{2}}{r} +
\frac{eBr}{2}\right)F + \left(E + m + \frac{Z\alpha}{r}\right)G =
0~,
\label{d1} \\
\frac{dG}{dr} + \left(\frac{l+\frac{1}{2}}{r} +
\frac{eBr}{2}\right)G - \left( E - m + \frac{Z\alpha}{r}\right)F
= 0~, \label{d2}
\end{eqnarray}
where $\alpha\equiv e^2=1/137$ is the fine structure constant. In
a strong magnetic field the asymptotic solutions of $F(r)$ and
$G(r)$ have the forms $\exp(-eBr^2/4)$ at large $r$, and
$r^\gamma$ with $\gamma = \sqrt{(l+1/2)^2 - (Z\alpha)^2}$ for
small $r$. One must have $Z\alpha <1/2$, otherwise the wave
function will oscillate as $r\to 0$ when $l=0$ and $l=-1$. With
these asymptotic factors, we now write
\begin{eqnarray}
F(r)=r^\gamma\exp(-eBr^2/4)~Q(r),~~~~
G(r)=r^\gamma\exp(-eBr^2/4)~P(r)~. \label{d3}
\end{eqnarray}
Substituting  Eq.(\ref{d3}) into Eq.(\ref{d1}) and (\ref{d2}) and
eliminating $P(r)$ from the coupled equations, we obtain
\begin{eqnarray}
\left\{\frac{d^2}{dr^2}+\left[\frac{2\gamma}{r}-eBr+\frac{Z\alpha/r^2}
{E+m+Z\alpha/r}\right]\frac{d}{dr}
+E^2-m^2\right.~~~~~~~~\nonumber\\
+\frac{2EZ\alpha}{r}+\frac{l+\frac{1}{2}}{r^2}-\frac{\gamma}{r^2}-eB(\Gamma+
1 ) ~~~~~~~~~~~~~~\nonumber\\ \left.
+\frac{Z\alpha/r^2}{E+m+Z\alpha/r}\left[\frac{\gamma}{r}-eBr-\frac{l+1/
2 } {r}\right]\right\}~Q(r)=0~, \label{d4}
\end{eqnarray}
where $\Gamma = l + 1/2 + \gamma$. Once $Q(r)$ is solved, the
form of $P(r)$ is obtainable from Eqs.(\ref{d1}) and (\ref{d3}).
If we let $x=r/l_B$, $l_B=1/\sqrt{eB}$, Eq.(\ref{d4}) becomes
\begin{eqnarray}
\left\{\frac{d^2}{dx^2}+\left[\frac{2\gamma}{x}-x+\frac{Z\alpha}
{x((E+m)l_Bx+Z\alpha)}\right]\frac{d}{dx}\right.~~~~~~~~~~~~~~~\nonumber\\
+(E^2-m^2)l_B^2+\frac{2Ezl_B\alpha}{x}+\frac{(l+1/2-\gamma)}{x^2}
-(\Gamma+1)~~~~\nonumber\\
\left.-\frac{Z\alpha(l+1/2-\gamma)}{x^2\left[(E+m)l_Bx+Z\alpha\right]}
-\frac{Z\alpha}{(E+m)l_Bx+Z\alpha}\right\}~Q(x)=0~. \label{d5}
\end{eqnarray}
Eq.(\ref{d5}) can be rewritten as
\begin{eqnarray}
\left\{\frac{d^2}{dx^2}+\left[\frac{2\beta}{x}-x-\frac{1}{x+x_0}\right]
\frac{d}{dx}+\epsilon+\frac{b}{x}-\frac{c}{x+x_0}\right\}~Q(x)=0~.
\label{d6}
\end{eqnarray}
Here $\beta=\gamma+1/2$, $x_0=Z\alpha/[(E+m)l_B]$,
$\epsilon=(E^2-m^2) l_B^2-(\Gamma+1)$, $b=b_0+L/x_0$,
$b_0=2EZ\alpha l_B$, $L=(l+1/2-\gamma)$,
 and $c=x_0+L/x_0$.  The energy $E$ is determined once  the
values of $\epsilon$ and $x_0$ are known.  The corresponding value
of the magnetic field $B$ is then obtainable from the expression
$l_B=Z\alpha/[(E+m)x_0]$.  Solution of $\epsilon$ and $x_0$ is
achieved in \cite{ChHo1} by means of the Bethe ansatz equations.
There it was shown that Eq.(\ref{d6}) is QES when $\epsilon$ is a
non-negative integer, i.e. $\epsilon=n, ~~n=0,1,2\ldots$.  In this
case, the function $Q(x)$ is a polynomial of degree $n$.

If we eliminate $Q(x)$ from Eqs.(\ref{d1}) and (\ref{d2}) instead,
we will obtain a second differential equation of $P(x)$:
\begin{eqnarray}
\left\{\frac{d^2}{dx^2}+\left[\frac{2\beta}{x}-x-\frac{1}{x+x_0^\prime}\right]
\frac{d}{dx}+\epsilon^\prime+\frac{b^\prime}{x}-\frac{c^\prime}{x+x_0^\prime}
\right\}~P(x) = 0 ~, \label{d6-1}
 \end{eqnarray}
with $x_0^\prime=Z\alpha/[(E-m)l_B]$, $\epsilon^\prime=\epsilon
+1$, $b^\prime=b_0+c^\prime$, and $c^\prime=-\Gamma/x_0^\prime$.
Other parameters are as defined previously.  It is obvious that
Eq.(\ref{d6-1}) is in the same form as Eq.(\ref{d6}), and hence is
also QES.  In \cite{ChHo1} it was shown that Eq.(\ref{d6}) and
(\ref{d6-1}) gave the same QES spectrum when
\begin{eqnarray}
\epsilon=n~~,&&\epsilon^\prime=n+1~,~~n=0,1,2,\ldots~,\nonumber\\
b^\prime-c^\prime&=& b-c+x_0~,~\\ x_0^\prime x_0&=&
\frac{(Z\alpha)^2}{\Gamma +n+1}~.\nonumber
\end{eqnarray}
  The result that $\epsilon=n$ and $\epsilon^\prime =n+1$ implies that the
degree of the polynomial $P(x)$ is of one order higher than that
of $Q(x)$, i.e. when $\epsilon=n$, $Q(x)$ and $P(x)$ are of degree
$n$ and $n+1$, respectively.

In \cite{ChHo1} it was also shown that the quasi-exact solvability
of Eqs.(\ref{d6}) and (\ref{d6-1}) is not due to the $sl(2)$
Lie-algebra, which is responsible for the quasi-exact solvability
of most one-dimensional and effectively one-dimensional systems.
We now want to show that the relevant algebra is indeed
$osp(2,2)$. In what follows, we shall present a differential
representation of this algebra, and then use it to demonstrate how
$osp(2,2)$ underlies the QES structure of our Dirac system.

\vskip 0.5 truecm

\section{A Differential Representation of $osp(2,2)$}

The superalgebra $osp(2,2)$ is characterized by four bosonic
generators $T^{\pm,0}$, $J$ and four fermionic generators
$Q_{1,2}$, $\bar{Q}_{1,2}$.  These generators satisfy the
commutation and anti-commutation relations \cite{ST,Tur3}
\begin{eqnarray}
[T^0,T^\pm]=\pm~T^\pm~~,~~~[T^+,T^-]=-2T^0~~,~~~[J,T^\alpha]=0~~,~~~\alpha=0,
+, -~~,\nonumber
\end{eqnarray}
\begin{eqnarray}
\{Q_1,\bar{Q_2}\}=-T^-~~,~~~\{Q_2,\bar{Q_1}\}=T^+~~,~~~\nonumber
\end{eqnarray}
\begin{eqnarray}
\frac{1}{2}(\{\bar{Q_1},Q_1\}+\{\bar{Q_2},Q_2\})=J~~,~~~
\frac{1}{2}(\{\bar{Q_1},Q_1\}-\{\bar{Q_2},Q_2\})=T^0~~,~~~\nonumber
\end{eqnarray}
\begin{eqnarray}
[Q_1,T^+]=Q_2~~,~~~[Q_2,T^+]=0~~,~~~[Q_1,T^-]=0~~,~~~[Q_2,T^-]=-Q_1~~,\label{osp}
\end{eqnarray}
\begin{eqnarray}
[\bar{Q}_1,T^+]=0~~,~~~[\bar{Q}_2,T^+]=\bar{Q}_1~~,~~~[\bar{Q}_1,T^-]
=\bar{Q}_2~~,~~~[\bar{Q}_2,T^-]=0~~,\nonumber
\end{eqnarray}
\begin{eqnarray}
[Q_{1,2},T^0]=\pm\frac{1}{2}Q_{1,2}~~,~~~[\bar{Q}_{1,2},T^0]
=\mp\frac{1}{2}\bar{Q}_{1,2}~~,~~~\nonumber
\end{eqnarray}
\begin{eqnarray}
[Q_{1,2},J]=-\frac{1}{2}Q_{1,2}~~,~~~[\bar{Q}_{1,2},J]=\frac{1}{2}\bar{Q}_{1,2}~~.\nonumber
\end{eqnarray}

A differential representation of the algebra $osp(2,2)$ can be
realized by the following $2\times 2$ differential-matrix
operators:
\begin{eqnarray}
T^+_n= \left(
\begin{array}{cc}
x^2d_x-nx&0\\ 0&x^2d_x-(n+1)x
\end{array}
\right)~~,~~~\nonumber
\end{eqnarray}
\begin{eqnarray}
T^0_n= \left(
\begin{array}{cc}
xd_x-\frac{n}{2}&0\\ 0&xd_x-\frac{n+1}{2}
\end{array}
\right)~~,~~~\nonumber
\end{eqnarray}
\begin{eqnarray}
T^-_n= \left(
\begin{array}{cc}
d_x&0\\ 0&d_x
\end{array}
\right)~~,~~~ J_n= \left(
\begin{array}{cc}
-\frac{n+2}{2}&0\\ 0&-\frac{n+1}{2}
\end{array}
\right)~~,~~~\label{osp-rep}
\end{eqnarray}
\begin{eqnarray}
Q_1= \left(
\begin{array}{cc}
0&0\\ 1&0
\end{array}
\right)~~,~~~\nonumber Q_2= \left(
\begin{array}{cc}
0&0\\ x&0
\end{array}
\right)~~,~~~\nonumber
\end{eqnarray}
\begin{eqnarray}
\bar{Q}_1= \left(
\begin{array}{cc}
0&xd_x-(n+1)\\ 0&0
\end{array}
\right)~~,~~~\nonumber \bar{Q}_2= \left(
\begin{array}{cc}
0&-d_x\\ 0&0
\end{array}
\right)~~.~~~\label{gen}
\end{eqnarray}
Here $x$ is a real variable, $d_x\equiv d/dx$, and $n$ is a real
number. It is easily checked that these matrices satisfy the
$osp(2,2)$ (anti-) commutator relations Eq.(\ref{osp}). For
non-negative integer $n$ ,  there exists for the $osp(2,2)$
algebra a $(2n+3)$-dimensional representation, with an invariant
subspace $P^n_{n+1}$ consisting of two-component functions of the
form
\begin{eqnarray}
\psi(x)= \left(
\begin{array}{c}
q_n(x)\\ p_{n+1}(x)
\end{array}
\right)~~,
\end{eqnarray}
where $q_n(x)$ and $p_{n+1}(x)$ are polynomials of degree $n$ and
$n+1$, respectively.

\vskip 0.5 truecm

\section{Hidden Lie-Algebraic Structure of The Dirac Equation }

We now show that the $osp(2,2)$ superalgebra is indeed the hidden
algebraic structure underlying the quasi-exact solvability of
Eq.(\ref{d6}) when $\epsilon$ is a non-negative integer $n\geq 0$.
First we rewrite Eq.(\ref{d6}) as
\begin{eqnarray}
T_Q(x) ~Q_n(x)&=&0~,\nonumber\\ T_Q(x)\equiv
(x^2+x_0x)\frac{d^2}{dx^2}&+&\left(-x^3-x_0x^2+2\beta~x+2\beta~x_0\right)
\frac{d}{dx}\nonumber\\ &+& nx^2+(nx_0+b-c)x+bx_0~.
\end{eqnarray}
We have put $\epsilon=n$, and write $Q(x)$ as $Q_n(x)$ to indicate
that it is a polynomial of degree $n$. This equation may be cast
into the matrix form
\begin{eqnarray}
{\bf T}_Q~\left(
\begin{array}{c}
Q_n(x)\\ 0
\end{array}
\right)\equiv
 \left(
\begin{array}{cc}
0&0\\ T_Q(x)&0
\end{array}
\right)
\left(
\begin{array}{c}
Q_n(x)\\ 0
\end{array}
\right)=0 ~~.~~~\label{mat-1}
\end{eqnarray}
The operator ${\bf T}_Q (x)$ is the $2\times 2$ matrix operator in
Eq.(\ref{mat-1}). Written in this form, the superalgebraic
structure hidden in it can be clearly exhibited. The $2\times 2$
matrix operator ${\bf T}_Q$ turns out to be expressible as a
linear combination of the $osp(2,2)$ generators in
Eq.(\ref{osp-rep}).  The two-component wave function in
Eq.(\ref{mat-1}) is simply an element in the invariant subspace
$P^n_{n+1}$ of the algebra with the lower component $p_{n+1}(x)=0$
(the form of $p_{n+1}(x)$ is immaterial here, as ${\bf T}_Q$
annihilates the lower component of any element in the subspace
$P^n_{n+1}$) . We shall demonstrate this below.

First let us express the various terms in ${\bf T}_Q$ in terms of
the generators in Eq.(\ref{osp-rep}).  After some algebras, we
obtain the following results:
\begin{eqnarray}
2Q_2T^0_nT^-_n-Q_1T^+_nT^-_n= \left(
\begin{array}{cc}
0&0\\ x^2d^2_x&0
\end{array}
\right)~~,~~~\nonumber
\end{eqnarray}

\begin{eqnarray}
Q_2T^-_nT^-_n= \left(
\begin{array}{cc}
0&0\\ xd^2_x&0
\end{array}
\right)~~,~~~~ Q_2T^+_n= \left(
\begin{array}{cc}
0&0\\ x^3d_x-nx^2&0
\end{array}
\right)~~,~~~\nonumber
\end{eqnarray}

\begin{eqnarray}
2Q_2T^0_n-Q_1T^+_n= \left(
\begin{array}{cc}
0&0\\ x^2d_x&0
\end{array}
\right)~~,~~~\nonumber
\end{eqnarray}

\begin{eqnarray}
2(Q_2T^0_n-Q_1T^+_n)= \left(
\begin{array}{cc}
0&0\\
nx&0
\end{array}
\right)~~,~~~\nonumber
\end{eqnarray}

\begin{eqnarray}
Q_2T^-_n= \left(
\begin{array}{cc}
0&0\\ xd_x&0
\end{array}
\right)~~,~~~~ Q_1T^-_n= \left(
\begin{array}{cc}
0&0\\ d_x&0
\end{array}
\right)~~.~~~\nonumber
\end{eqnarray}

With these expressions, the operator ${\bf T}_Q$ can then be
written as
\begin{eqnarray}
{\bf T}_Q=(2Q_2T^0_nT^-_n-Q_1T^+_nT^-_n)
+x_0Q_2T^-_nT^-_n-Q_2T^+_n~\nonumber\\
-x_0(2Q_2T^0_n-Q_1T^+_n)+2\beta(Q_2T^-_n+x_0Q_1T^-_n)~\nonumber\\
 +2x_0(Q_2T^0_n-Q_1T^+_n)+(b-c)Q_2+bx_0Q_1~~.
\end{eqnarray}
We have therefore succeeded in expressing ${\bf T}_Q$ as a linear
combination of the generators of $osp(2,2)$ with finite
dimensional subspace.  The underlying algebraic structure
responsible for the quasi-exact solvability of Eq.(\ref{d6}) is
thus demonstrated.

The underlying algebraic structure responsible for the quasi-exact
solvability of Eq.(\ref{d6-1}) can be obtained in the same way. In
fact, as mentioned in Sect.~2, Eq.(\ref{d6-1}) is in the same form
as Eq.(\ref{d6}), only with the parameters $x,~b,~c$ and $n$ being
replaced by $x^\prime,~b^\prime,~c^\prime$ and $n+1$,
respectively. Hence, we immediately see that the corresponding
operator ${\bf T}_P$ will be of exactly the same form as ${\bf
T}_Q$ with the same replacements of the corresponding parameters.

\section{Summary}
In this paper we have unveiled the underlying symmetry responsible
for the quasi-exact solvability of planar Dirac electron in a
Coulomb and a magnetic field.  The relevant second order
differential operator acting on any one component of the
two-component wave function was recast in a $2\times 2$ matrix
form.  This $2\times 2$ differential-matrix operator was then
shown to be expressible as a linear combination of the generators
of the superalgebra $osp(2,2)$, thus exhibiting the algebraic
structure of the QES Dirac system.  With this result, all the
algebraic structures making the systems of
 planar charged particles in
Coulomb and magnetic fields QES have been identified.

\vskip 2cm \centerline{\bf Acknowledgment}

This work was supported in part by the National Science Council of
the Republic of China through Grant No. NSC 93-2112-M-032-009.

\end{document}